# ANALYSIS OF SPEECH UNDER STRESS USING LINEAR TECHNIQUES AND NON-LINEAR TECHNIQUES FOR EMOTION RECOGNITION SYSTEM

A. A. Khulage and Prof. B. V. Pathak.

M.E student, Cummins college of Engineering, Pune.

`asmita_152007@rediffmail.com`
Assistant professor, Cummins college of Engineering, Pune.

`bvpathak100@yahoo.com`

## ABSTRACT

*Analysis of speech for recognition of stress is important for identification of emotional state of person. This can be done using 'Linear Techniques', which has different parameters like pitch, vocal tract spectrum, formant frequencies, Duration, MFCC etc. which are used for extraction of features from speech. TEO-CB-Auto-Env is the method which is non-linear method of features extraction. Analysis is done using TU-Berlin (Technical University of Berlin) German database. Here emotion recognition is done for different emotions like neutral, happy, disgust, sad, boredom and anger. Emotion recognition is used in lie detector, database access systems, and in military for recognition of soldiers' emotion identification during the war*.

## 1. INTRODUCTION

A speech signal is introduced into a medium by a vibrating object as vocal folds in throat. Thich is the source of the disturbance that moves through the medium. Each spoken word is created using the phonetic combination of a set of vowel semivowel and consonant speech sound units. Different stress is applied by vocal cord of a person for particular emotion. Speech signal representation Matlab is shown in figure 1.

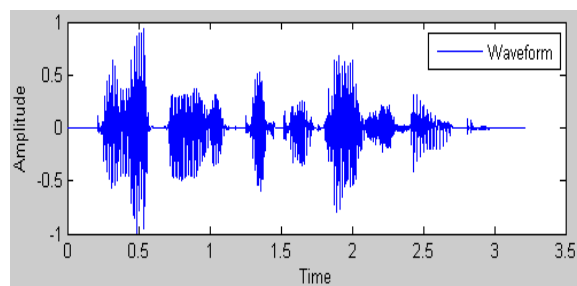

Figure 1. Speech signal representation.

Stress is a psycho-physiological state which is characterized by subjective strain, dysfunctional physiological activity of the speech signal. Stress may be induced by external factors such as workload, noise, vibration, sleeps loss, etc. and by internal factors like emotion, fatigue, etc. Physiological consequences of stress are respiratory changes. This respiratory change can be increased respiration rate, irregular breathing, increased muscle tension of the vocal cords, etc. The increased muscle tension of the vocal cords and vocal tract can directly or indirectly and adversely affect the quality of speech. Various feature extraction techniques are implemented for speech recognition which is basically classified as linear techniques and non-linear techniques. Formant frequencies estimation, pitch frequencies, duration, MFCC, vocal tract spectrum fall under linear techniques and TEO-CB-Auto is a non-linear technique. These techniques are used for emotion recognition.

Emotion identification is used in different applications such as Lie detector and can be used as voice tag in different database access systems. This voice tag is used in telephony shopping, ATM machine as a password for accessing that particular account.

Formants are resonances of vocal tract and estimation of their location and frequencies at that location which is important in emotion recognition. Formant frequencies estimation Technique involves extraction of resonance peaks from filter coefficients obtained through LPC analysis of a speech waveform. After the calculation of prediction polynomial A(z), formants parameters are determined by " peak picking".

Duration parameter is calculated using extraction of vowel, semivowel and consonant duration in that speech signals for emotion recognition.

Vocal tract spectrum estimation or analysis is considered on the basis of bandwidth calculations of speech signal in frequency domain.

MFCCs are coefficients which represent audio, based on perception of human auditory systems. The basic difference between the operation of FFT/DCT and the MFCC is that in the MFCC, the frequency bands are positioned logarithmically (on the mel scale) which approximates the human auditory system's response more closely than the linearly spaced frequency bands of FFT or DCT.

Non-linear based TEO-CB-Auto-Env method is based on the principle that the true source of sound production is actually the vortex-flow interactions, which are nonlinear in nature. Changes in vocal system physiology induced by speech spoken under stressful conditions such as muscle tension will affect the vortex-flow interaction patterns in the vocal tract.

Based on the theory that hearing could be viewed as the process of detecting the energy, TEO-CB-Auto-Env method is developed. This technique is very useful for emotion recognition.

## 2. FEATURE EXTRACTION

When we pronounce any particular alphabet or word in different styles, vocal tract produces them by varying its' dimension. Emotions are expressed by applying a specific amount of stress on that word or alphabet. For recognition of style (it may be sad, angry, disgust, boredom, neutral or happy), different feature extraction techniques are used [3].

### 2.1. Linear Techniques

The parameters generally considered in evaluating changes in speech produced under influence of perceptual and physical stress (noise, vibration, and g-force) are: intensity, pitch, duration, vocal tract spectrum, glottal source and vocal tract articulators' profiles [3][4][6].

### 2.1.1 Formant Frequencies

Formants are nothing but the spectral peaks of the sound spectrum |P(f)| of the voice. In speech science and phonetics, formant frequencies are an acoustic resonance of the human vocal tract which is measured as an amplitude peak in the frequency spectrum of the sound.

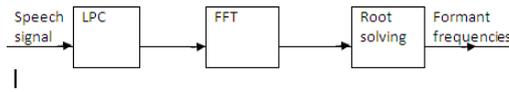

Figure 2 Block diagram of formant frequency detection using LPC.

In acoustics, formants are referred as a peak in the sound envelope and/or to a resonance in sound sources, as well as that of sound chambers. Extraction of Formant Frequencies is done using LPC Based Formants Estimation Technique. The vocal tract is modeled as a linear filter with resonances and resonance frequencies of the vocal tract are called formant frequencies [2]. Graphically, the peaks of the vocal tract response of speech signal correspond roughly to its formant frequencies. If the vocal tract is modeled as a time-invariant, all-pole linear system, then each of the conjugate pair of poles that corresponds to a formant frequency or resonance frequency [2]. Graphical representation of Formant frequencies estimation for a speech signal in neutral emotional state is shown in figure 3.

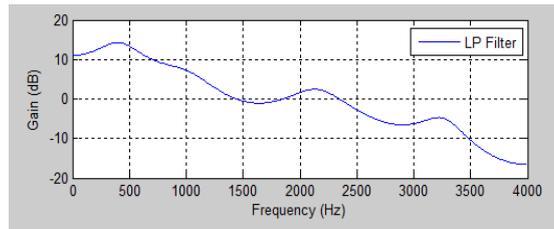

Figure 3. Graphical representation of formant frequencies for angry emotion

Here, the review of mechanics of computing a linear prediction model, and after that the implications in formant frequencies estimation is explained.

In fact, the speech signal can be defined as:

$$S(n) = -\sum_{i=1}^{N_{LP}} a_{LP}(i) \cdot s(n-i) + e(n) \quad (1)$$

Where, $N_{LP}$ and $a_{LP}$ represents, respectively, the number of coefficients in the model, the linear prediction coefficients. $e(n)$ is the error in the model. LPC analysis of speech signal produces predictor polynomial of degree at least 10 which, due to stability requirement, has all its roots within unit circle [2]. Equation (1) is written in Z-transform notation as a linear filtering operation,

$$E(z) = H_{LP}(z) \cdot S(z) \quad (2)$$

E(z) and S(z) is denoted, respectively, the Z-transform of the error signal and the speech signal. HLP(z) is a linear prediction inverse filter.

$$H_{LP}(z) = \sum_{i=0}^{N_{LP}} a_{LP}(i) \cdot z^{-i} \quad (3)$$

Formant frequencies can be estimated from the LP smoothed spectrum and using spectrum, local maxima are found and those of small bandwidths are related to formants [3]. After that, Peak-picking can be used to estimate formants. [2]. Estimation of formant frequencies based on the relationship between formant and poles of the vocal tract filter is considered [2].

The denominator of the transfer function may be factored,

$$1 + \sum_{i=0}^{N_{LP}} a_{LP}(i) \cdot z^{-i} = \prod_{k=0}^{N_{LP}} (1 - c_k \cdot z^{-1})$$

Where, $C_k$ are a set of complex numbers. Each complex conjugate pair of poles representing a resonance at frequency:
$F_k = (F_s/2*pi) \tan^{-1}[Im(c_k)/Re(c_k)]$
And bandwidth:
$B_k = -(F_s/pi) * \ln(c_k)$
If the pole lies close to the unit circle then the root represents a formant frequency.
$r_k = (Im(c_k)^2 + Re(c_k)^2))^{1/2} \geq 0.7$

### 2.1.2 Vocal Tract Spectrum

Analysis of vocal-tract spectrum is based on formant location and bandwidth for selected phonemes across the selected database of speech signals [3].

Statistical evaluations such as mean and variance estimates for specific phonemes were analyzed for different emotions. Statistical evaluations showed that happy emotion of speakers was statistically different from neutral, sad and boredom [3][4]. Vocal tract information for happy were the most consistent across the bandwidths tested. A snapshot of a matlab command window for vocal tract spectrum for neutral emotional state is shown in figure 4.

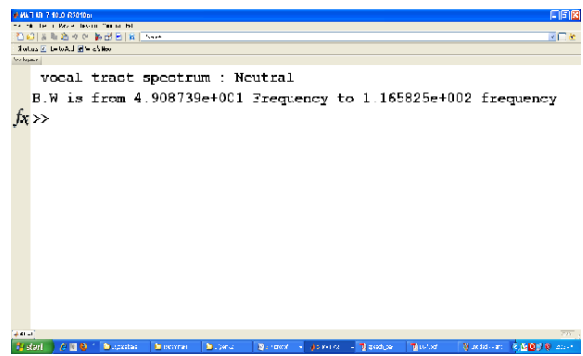

figure 4: Snapshot of matlab command window showing the bandwidth range for neutral emotion.

For Bandwidth calculation, speech signal in time domain is converted to frequency domain. Minimum and maximum value of frequencies is chosen from array of frequencies.
Formula for bandwidth is given by,
BW= Fmax - Fmin

### 2.1.3 Speech Duration

A speech duration study of speech under stress considers statistical evaluations of individual Phoneme class duration. Duration analysis was conducted separately across the speech classes such as vowel, consonant, and semivowel. An analysis was also conducted across interclass duration movement to determine if speakers increased duration of certain phoneme classes for particular emotion at the expense of others. Examples of overall word phone class (vowel, semivowel, and consonant) duration for neutral and disgust emotions.

Speech signal contains vowels in first three formant frequencies where as semivowel extraction is done using calculation of frequency difference between $1^{st}$ and $2^{nd}$ formant frequencies, difference between $3^{st}$ and $2^{nd}$ formant frequencies, difference between $4^{th}$ and $2^{nd}$ formant frequencies of speech signal.

Consonants are extracted using energy between 0-300Hz and 640-2800Hz of speech signal of formant frequencies waveform.

Mean word duration was a significant indicator for speech in disgust when compared to neutral

[3][4].

Vowel duration is highest as compared to semivowel and consonant duration. A snapshot of matlab command window for duration calculation of neutral emotion is shown in figure 5.

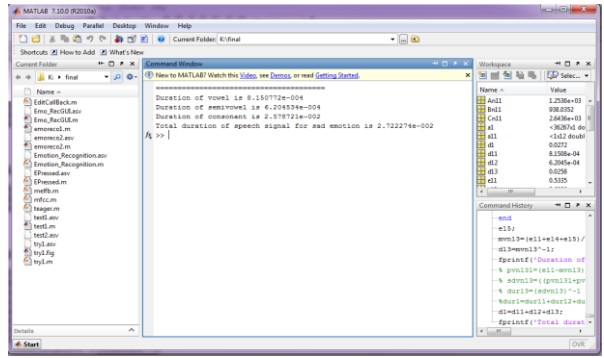

Figure 5. Matlab command window for duration calculation of sad emotional state speech signal

### 2.1.4 MFCC

MFCCs are calculated from the Discrete Cosine Transform (DCT) of the audio clip. The MFCC process is carried out by five phases as shown in fig 6. In the frame blocking section, the speech waveform is divided into frames of approximately 30 milliseconds. Windowing minimizes the discontinuities present in the signal by tapering the beginning and end of each frame to zero. The FFT is used for conversion of each frame from the time domain to the frequency domain. Considering the Mel frequency wrapping operation, the signal is plotted against the Mel-spectrum to mimic human hearing. According to speech science, human hearing does not follow the linear scale but rather the Mel-spectrum scale which is a linear spacing below 1000 Hz and logarithmic scaling above 1000 Hz. Finally, the Mel-spectrum plot is converted back to the time domain by using the following equation given below,

Mel (f1) = 2595*log10 (1 + f1 /700)

The resultant matrices are the Mel-Frequency Cepstrum Coefficients.

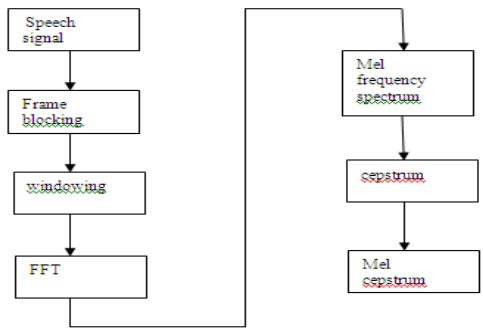

Figure 6 MFCC Block Diagram

Figure 7 shows matlab command window showing mean of MFCC values for neutral emotion.

Figure 7 : Matlab command window snapshot for neutral emotion

## 2.2. Non-Linear Techniques

All speech features used in linear techniques are derived from a linear speech production models which assume that airflow propagates in the vocal tract as a plane wave. This pulsatile flow can be the source of sound production. This assumption may not give correct results since the flow is actually separate and concomitant vortices are distributed throughout the vocal tract.

Non-Linear method is based on principle that the source of sound production is actually the vortex-flow interactions, which are nonlinear. TEO based Linear technique has three sub-types namely TEO-FM-Var, TEO-Auto-Env and TEO-CB-Auto-Env. TEO-CB-Auto-Env method is used for feature extraction in this paper because of more accuracy.

Hearing can be a process of detecting the energy of speech signal. Based on the principle of the oscillation pattern of a simple spring mass system, TEO-CB-Auto-Env developed an energy operator to measure the energy for simple sinusoids which is useful elements for speech [7].

### 2.2.1 TEO-CB-Auto-Env

To determine the TEO-CB-Auto-Env feature, each TEO profile output is segmented into 200-sample (25 msec) frames. Normalized TEO autocorrelation envelope area parameters are extracted of speech signal. Final output of TEO-CB-Auto-Env is considered as the area under envelope of TEO autocorrelation. Block diagram for TEO-CB-Auto-Env is shown in figure 8. TEO-CB-Auto-Env method is having highest accuracy as compared to other two methods of TEO profile.

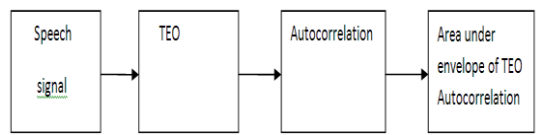

Figure 8: TEO-CB-Auto-Env Block Diagram

Teager Energy Operator for discrete-time signals from its discrete form is given as,
$y(n) = abs(x(n))^2 - x(n+1) \ast conj(x(n-1))$
Where $x(n)$ is the sampled speech signal,
$x(n+1)$ and $x(n-1)$ is shifted signal of original speech signal $x(n)$.
Autocorrelation of speech signal is given by,
$R_{xx}(j) = \sum_n x_n \cdot x_{n-j}$

TEO-CB-Auto-CB envelope and its calculated energy is respectively shown in figure 9 and 10.

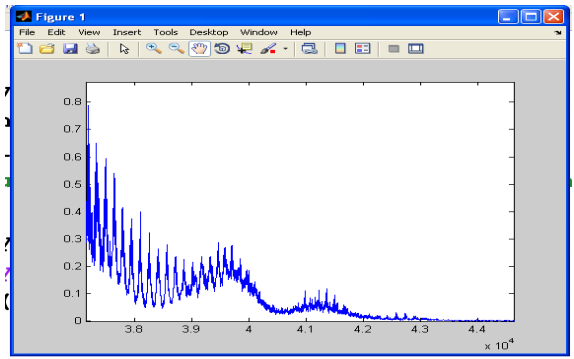

Figure 9: TEO-Autocorrelation Envelope

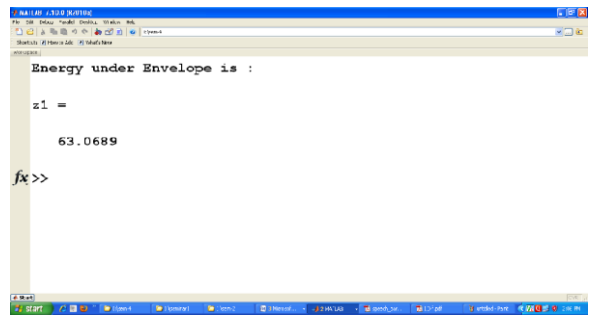

Figure 10: Calculated energy under envelope

## 3. EMOTION RECOGNITION

The idea of the audio signal processing is to implement a recognizer using Matlab which can identify person's emotional state by processing his/her voice. The Matlab functions and scripts were all well documented and parameterized in order to identify emotional state. The basic goal of our project is to recognize and classify the emotions of speech signal. This is done using hierarchical algorithm. Classification using hierarchical algorithm is mainly based on extracting several key features like duration, vocal tract spectrum estimation, MFCC, formant frequencies etc from the speech signals of those persons by using the process of feature extraction using Matlab. Emotion recognition process using different feature extraction techniques is explained below which differentiates six emotions such as neutral, angry, happy, sad, disgust and boredom etc of a speech signal.

First of all, speech signal is passed through TEO-CB-Auto-Env technique. The energy of envelope of signal was used to classify emotional state, which separate out angry and disgust emotion from neutral, happy, sad and boredom. Using formant frequencies estimation angry and disgust emotions are recognized. Out of sad, happy, disgust and boredom emotions, happy emotion is distinguished using vocal tract spectrum estimation. Sad emotion is recognized using duration calculation of a speech signal. Mean of MFCC values are used for identification of neutral and boredom emotion. Recognition procedure for angry emotion is shown in figure 11.

figure 11. Recognition procedure of speech signal for angry emotional speech

## 4. DATABASE OF GERMAN EMOTIONAL SPEECH

Database of speech for different emotions is created by the institute of communication science of the TU-Berlin (Technical University of Berlin) to examine acoustical correlates of emotional speech. An emotional database comprising 6 basic emotions (anger, joy, sadness, fear, disgust and boredom) as well as neutral speech.

Ten professional native German actors (5 female and 5 male) simulated these emotions, producing 10 utterances (5 short and 5 longer sentences), which could be used in every-day communication and are interpretable in all applied emotions.

The recorded speech material of about 800 sentences (7 emotions * 10 actors * 10 sentences + some second versions) are now getting evaluated with respect to recognizability and naturalness in a forced-choice automated listening-test by 20-30 judges.
Those utterances for which the emotion was recognized by at least 80 % of the listeners will be used for further analysis.

On the basis of this database analysis of distinguishable features of the specific emotion is done for emotion recognition system. The perceptive relevance of typical features of the emotions is getting evaluated by means of speech-re-synthesis, allowing the controlled variation of specific features.

## 5. RESULT

Analysis of speech signal for emotion (neutral, angry, happiness, sad, disgust and boredom) is done using hierarchical algorithm that uses different feature extraction techniques. TEO-CB-Auto-Env method is used as basic classification technique for different emotions for speech signal.

The energy of envelope of signal was used to classify disgust or angry emotional state. Using formant frequencies estimation angry and disgust emotions are recognized. Vocal tract spectrum estimation is used for identification of happy emotion. Duration calculation is used for identification of sad emotion. Boredom and neutral emotions are classified using mean of MFCC values. Recognition procedure for speech signal in neutral emotional state is given in figure 12.

figure 12. Neutral speech recognition procedure.

## 6. CONCLUSION

Hierarchical algorithm is used for classification of different emotions. After differentiating anger and disgust emotions from TEO-CB-Auto method, formant frequencies estimation method is used to classify angry and disgust emotion recognition. First formant frequencies range is has higher values in angry emotion as compared to disgust emotion. Vocal tract spectrum estimation is used for identification of happy emotion. Sad emotion is separate out from boredom and neutral emotion using duration calculation. Boredom and neutral emotion is differentiated under the consideration of mean values of MFCC. Mean values of MFCC in boredom emotion are greater than mean values of MFCC obtained in case of neutral emotion.

## ACKNOWLEDGEMENT

The authors would like to thank the reviewers for constructive suggestions.